\documentclass[aps, prd, groupedaddress, preprint, tightenlines,  eqsecnum, showpacs, nofootinbib]{revtex4}

\usepackage{natbib}

\usepackage{epsfig}
\usepackage{graphicx}

\usepackage{axodraw, float, slashed, graphicx, amssymb, amsmath}
\usepackage[usenames, dvipsnames]{xcolor}

\def\Li{\mathop{\hbox{\rm Li}}\nolimits}

\def\ksl{\slashed{k}}

\def\Tr{{\rm Tr}}

\def\Fcc{P^{(2)}}
\def\ki{a}
\def\kj{b}

\def\spa#1.#2{\left\langle#1\,#2\right\rangle}
\def\spb#1.#2{\left[#1\,#2\right]}
\def\la{\langle}
\def\ra{\rangle}

\DeclareMathOperator{\tr}{\mathrm{Tr}}

\def\eps{\epsilon}

\def\be{\begin{equation}}
\def\ee{\end{equation}}

\begin{document}

\hfill\today

\title{Color Dressed Unitarity and Recursion for Yang-Mills Two-Loop All-Plus Amplitudes }

\author{David~C.~Dunbar, John~H.~Godwin, Warren~B.~Perkins and  Joseph~M.W.~Strong}

\affiliation{
College of Science, \\
Swansea University, \\
Swansea, SA2 8PP, UK\\
\today
}

\begin{abstract}
We present a direct computation of the full color two-loop five-point all-plus Yang-Mills amplitude 
using four dimensional unitarity and recursion. We present the $SU(N_c)$ amplitudes in compact analytic forms.
Our results match the explicit expressions previously computed but do not require full two-loop
integral methods. 
\end{abstract}


\maketitle

\section{Introduction}

Computing perturbative scattering amplitudes is not only a key tool in confronting theories of particle physics with experimental results but is also a 
gateway for exploring 
the symmetries and properties of theories which are not always manifest in a Lagrangian approach. 
Since the standard model of particle physics and many of its potential extensions are gauge 
theories, gauge theory amplitudes are of particular interest.  
Within a Yang-Mills gauge theory a $n$-gluon amplitude in   may be expanded in the gauge coupling constant,
\begin{equation}
{\cal A}_n = g^{n-2}  \sum_{\ell\geq 0} a^{\ell}{\cal A}_n^{(\ell)} 
\end{equation}
where $a=g^2e^{-\gamma_E \epsilon}/(4\pi)^{2-\epsilon}$.  Each loop amplitude can be
further expanded in terms of color structures, $C^\lambda$,
\begin{equation}
{\cal A}_n^{(\ell)} = \sum_{\lambda} { A}_{n:\lambda}^{(\ell)} C^\lambda\,,
\end{equation}
separating the color and kinematics of the amplitude. The color structures $C^\lambda$ may be organised in terms of powers of $N_c$.

A great deal of progress has been made in computing ${\cal A}_n^{(\ell)}$ for tree amplitudes ($\ell=0$) and one-loop amplitudes ($\ell=1$)
in $SU(N_c)$ gauge theory.
However progress in two-loop amplitudes has been more modest:
the four gluon amplitude 
has been computed \cite{Glover:2001af,Bern:2002tk} for the full color and helicity structure and there is currently 
tremendous progress in the computation of the five-point amplitude.  The first amplitude to be computed at five point was the leading in 
color part of the amplitude with all positive helicity external gluons (the all-plus amplitude) 
which was computed using $d$-dimensional  unitarity methods~\cite{Badger:2013gxa,Badger:2015lda}
and was subsequently presented in a very elegant and compact form~\cite{Gehrmann:2015bfy}. 
In~\cite{Dunbar:2016aux}, it was shown how four-dimensional unitarity techniques could be used to regenerate the five-point leading in color amplitude
and in \cite{Dunbar:2016gjb,Dunbar:2017nfy} the leading in color all-plus amplitudes were obtained for six- and seven-points,
these being the first six- and seven-point amplitudes to be obtained at two-loops. 
The leading in color five-point amplitudes have been computed for the remaining helicities~\cite{Abreu:2019odu,Badger:2018enw}.  
Full color amplitudes are significantly more complicated requiring a larger class of master integrals incorporating 
non-planar integrals~\cite{Chicherin:2018old,Chawdhry:2018awn}. In~\cite{Badger:2019djh} the first full color five-point amplitude was presented in QCD.

In this article we will examine the one and two-loop partial amplitudes using a $U(N_c)$ color trace basis where the fundamental objects are traces of color 
matrices $T^{a}$ rather than contractions of the structure constants $f^{abc}$. 
We  examine the particular scattering amplitude in pure gauge theory where the external gluons have identical 
helicity, ${\cal A}_n(1^+,\cdots n^+)$. 
This amplitude is fully crossing symmetric which makes computation relatively more tractable but nonetheless is a valuable laboratory for studying the 
properties of gluon scattering.
The all-plus amplitude has a singular structure which is known from general theorems together with a finite remainder part. 
We present a form for the finite part which is a simple combination of dilogarithms together with rational terms. 
Specifically we compute directly all the color trace structures for the five-point all-plus two-loop amplitude.  
Our results are in complete agreement with the results recently computed by Badger et. al.~\cite{Badger:2019djh} and are consistent with constraints 
imposed by group theoretical arguments~\cite{Naculich:2011ep,Edison:2011ta}.

Our methodology involves computing the polylogarithmic and rational parts of the finite remainder by a combination of techniques. 
The polylogarithms are computed using four dimensional unitarity cuts and the rational parts are determined by recursion.  
We use augmented recursion~\cite{Alston:2012xd} to overcome the issues associated with the presence of double poles.

\section{One-Loop Sub-leading Amplitudes} 

An $n$-point tree amplitude can be expanded in a color trace basis as
\begin{eqnarray}
{\cal A}_n^{(0)}(1,2,3,\cdots ,n)  &=& \sum_{S_n/Z_n}  \Tr[ T^{a_1} \cdots T^{a_n}] A_{n:1}^{(0)} (a_1,a_2,\cdots a_n).
\end{eqnarray}
This separates the color and kinematic structures.
The partial amplitudes $A_{n:1}^{(0)} (a_1,a_2,\cdots a_n)$ are cyclically symmetric but not fully crossing symmetric.  The sum over permutations is over
permutations of $(1,2,\cdots n)$ up to this cyclic symmetry.   This is not the only expansion and in fact other expansions exist~\cite{DelDuca:1999rs}
which  may be more efficient for some purposes.  This color decomposition is valid for both  $U(N_c)$ and $SU(N_c)$ gauge theories.  
If any of the external particles in the $U(N_c)$ case are $U(1)$ particles then the amplitude must vanish.  
This imposes {\it decoupling identities} amongst the partial amplitudes~\cite{Bern:1990ux}.  
For example at tree-level setting  leg $1$ to be $U(1)$ and extracting the coefficient of 
$\Tr[T^2 T^3 \cdots T^n]$ implies that
\begin{equation}
A_{n:1}^{(0)}(1,2,3,\cdots n ) +A_{n:1}^{(0)}(2,1,3,\cdots n )+\cdots A_{n:1}^{(0)}(2,\cdots, 1, n )=0.
\label{eq:decotree}
\end{equation}
This provides a consistency check on the partial amplitudes. At loop level these decoupling identities  provide powerful relationships between the
different pieces of the amplitude.

In a $U(N_c)$ gauge theory the one-loop $n$-point amplitude can be expanded as~\cite{Bern:1990ux}
\begin{eqnarray}
& & {\cal A}_n^{(1)}(1,2,3,\cdots ,n)  = \sum_{S_n/Z_n}   N_c \Tr[ T^{a_1} \cdots T^{a_n}] A_{n:1}^{(1)} (a_1,a_2,\cdots a_n)
\notag
\\
&+&\sum_{r=2}^{[n/2]+1} \sum_{S_n/(Z_{r-1}\times Z_{n+1-r})}
\hskip-1.0truecm
\Tr[ T^{a_1} \cdots T^{a_{r-1}}]\Tr[T^{b_r} 
\cdots T^{b_{n}}] A_{n:r}^{(1)}(a_1\cdots a_{r-1} ; b_r \cdots b_{n})\,.
\label{eq:oneloopcolordeco}
\end{eqnarray}
The $A_{n:2}^{(1)}$ are absent (or zero) in the $SU(N_c)$ case.
For $n$ even and $r=n/2+1$ there is an extra $Z_2$ in the summation to ensure each color structure only appears once.
The partial amplitudes  $A_{n:r}^{(1)}(a_1\cdots a_{r-1} ; b_r \cdots b_{n})$ are cyclically symmetric in the sets $\{a_1\cdots a_{r-1} \}$ and 
$\{b_r \cdots b_{n} \}$ and obey a ``flip'' symmetry,
\begin{equation}
A^{(1)}_{n:r} (1,2,\cdots (r-1) ; r, \cdots n) =(-1)^n A^{(1)}_{n:r} (r-1,\cdots 2,1 ; n, \cdots  r) \,.
\end{equation}  
Amplitudes involving the scattering of gauge bosons also occur in string theories.  From a string theory viewpoint the
$A_{n:r}^{(1)}$ with $r > 1$ would be considered non-planar contributions arising from attaching gauge bosons to the two edges of a one-loop surface. 
 
Decoupling identities impose relationships amongst the partial amplitudes.  For example setting  leg $1$ to be $U(1)$ and extracting the coefficient of 
$\Tr[T^2 T^3 \cdots T^n]$ implies
\begin{equation}
A_{n:2}^{(1)}(1;2,3,\cdots n) +A_{n:1}^{(1)}(1,2,3,\cdots n ) +A_{n:1}^{(1)}(2,1,3,\cdots n )+\cdots A_{n:1}^{(1)}(2,\cdots, 1, n )=0
\label{eq:decoupleA}
  \end{equation}
and consequently the $A_{n:2}^{(1)}$ can be expressed as a sum of $(n-1)$ of the $A_{n:1}^{(1)}$.  
By repeated application of the decoupling identities all the $A_{n:r}^{(1)}$ can be expressed as sums over the $A_{n:1}^{(1)}$~\cite{Bern:1990ux},
\begin{equation}
A_{n;r}^{(1)}(1,2,\ldots,r-1;r,r+1,\ldots,n)\ =\
 (-1)^{r-1} \sum_{\sigma\in COP\{\alpha\}\{\beta\}} A_{n;1}^{(1)}(\sigma)
\label{eq_COP}
\end{equation}
where $\alpha_i \in \{\alpha\} \equiv \{r-1,r-2,\ldots,2,1\}$ and
$\beta_i \in \{\beta\} \equiv \{r,r+1,\ldots,n-1,n\}$ [Note that the ordering of the first set of indices is reversed
with respect to the second].
$COP\{\alpha\}\{\beta\}$ is the set of all
permutations of $\{1,2,\ldots,n\}$ with $n$ held fixed
that preserve the cyclic
ordering of the $\alpha_i$ within $\{\alpha\}$ and of the $\beta_i$
within $\{\beta\}$, while allowing for all possible relative orderings
of the $\alpha_i$ with respect to the $\beta_i$.
For example if $\{\alpha\} = \{2,1\}$ and
$\{\beta\} = \{3,4,5\}$, then $COP\{\alpha\}\{\beta\}$
contains the twelve elements
\begin{eqnarray*}
(2,1,3,4,5),\quad (2,3,1,4,5),\quad (2,3,4,1,5),\quad
  (3,2,1,4,5),\quad (3,2,4,1,5),\quad (3,4,2,1,5), \\
 (1,2,3,4,5),\quad (1,3,2,4,5),\quad (1,3,4,2,5),\quad
  (3,1,2,4,5),\quad (3,1,4,2,5),\quad (3,4,1,2,5) \,.
\end{eqnarray*}

The simplest one-loop QCD $n$-gluon helicity amplitude is the {\it all-plus} amplitude  with all 
external helicities positive.  The tree amplitude vanishes for this particular amplitude and consequently, the one-loop amplitude is rational (to order $\epsilon^0$).
The leading in color one-loop partial amplitude has an all-$n$ 
expression~\cite{Bern:1993qk}
\footnote{Here  a null momentum is represented as a
pair of two component spinors $p^\mu =\sigma^\mu_{\alpha\dot\alpha}
\lambda^{\alpha}\bar\lambda^{\dot\alpha}$. 
We are using a spinor helicity formalism with the usual
spinor products  $\spa{a}.{b}=\epsilon_{\alpha\beta}
\lambda_a^\alpha \lambda_b^{\beta}$  and 
 $\spb{a}.{b}=-\epsilon_{\dot\alpha\dot\beta} \bar\lambda_a^{\dot\alpha} \bar\lambda_b^{\dot\beta}$. 
\noindent{Also}
 $ s_{ab}=(k_a+k_b)^2=\spa{a}.b \spb{b}.a=\la a|b|a]$,
$ {\rm tr}_-[ijkl]\equiv {\rm tr}( \frac{(1-\gamma_5)}{2} \ksl_{i} \ksl_{j} \ksl_{k} \ksl_{l} ) =\spa{i}.{j}\spb{j}.{k}\spa{k}.{l}\spb{l}.{i}$,\\
\noindent{
${\rm tr}_+[ijkl]\equiv {\rm tr}( \frac{(1+\gamma_5)}{2} \ksl_{i} \ksl_{j} \ksl_{k} \ksl_{l} ) =\spb{i}.{j}\spa{j}.{k}\spb{k}.{l}\spa{l}.{i}$
and $\epsilon(i,j,k,l)={\rm tr}_+[ijkl]-{\rm tr}_-[ijkl]$.}
}
\begin{equation}
A_{n:1}^{(1)}(1^+,2^+,\ldots,n^+)\ =\ -{i \over 3}\,
{ 1\over \spa1.2 \spa2.3 \cdots \spa{n}.1} { \sum_{1\leq i < j < k < l  \leq n}{\rm tr}_-[i j k l ]  } 
+ O(\eps) \,.
\end{equation}
This expression is order $\epsilon^0$ but all-$\epsilon$ expressions exist for the first few amplitudes in this series~\cite{Bern:1996ja}. 
The sub-leading terms can be obtained from decoupling identities. We have obtained compact expressions, to order $\epsilon^0$,  for these: 
\begin{align}
A_{n:1}^{(1)}(1^+,2^+,3^+,\cdots, n^+)  &= -\frac{i}{3}{1 \over \spa1.2\spa2.3\cdots \spa{n}.1}{ \sum_{1\leq i < j < k < l  \leq n}  
{\rm tr}_-[ijkl] }\,,
\notag\\
A_{n:2}^{(1)}(1^+ ;2^+,3^+,\cdots, n^+) &= -i{ 1 \over \spa2.3\spa3.4 \cdots \spa{n}.2} \sum_{2 \leq i<j \leq n}  \spb{1}.{i} \spa{i}.{j} \spb{j}.{1} 
\intertext{and for $r\ge3$}
A_{n:r}^{(1)}(1^+,2^+,\cdots, r-1^+ ;  r^+ \cdots n^+) &= -2i {  (K_{1\cdots r-1}^2)^2   \over 
(\spa1.2\spa2.3 \cdots \spa{(r-1)}.1)  ( \spa{r}.{(r+1)}  \cdots \spa{n}.{r}  )} \,.
\end{align}
These expressions are remarkably simple given the number of terms arising in the naive application of (\ref{eq_COP}):
the number of terms in the numerator of a single $A_{n:1}^{(1)}$ grows as
\begin{equation}
\frac{1}{24} n(n-1)(n-2)(n-3)
\end{equation}
while the summation over $COP$ terms grows with $n$ as
\begin{equation}
\sim \frac{(n-1)!}{(r-2)!(n-r)!}\,.
\end{equation}

\begin{figure}[H]
\begin{center}
\includegraphics{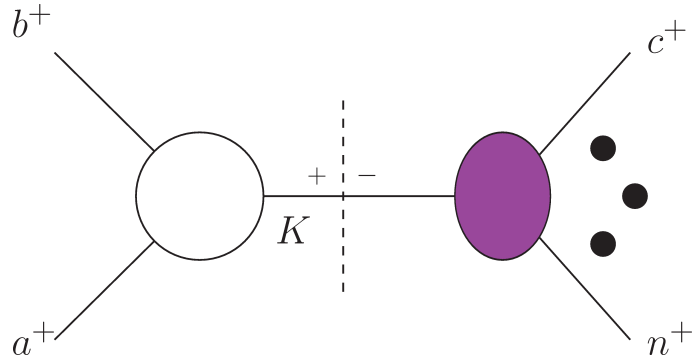}
\caption{The origin of the double pole. The double pole corresponds to the coincidence of the singularity arising in the 3-pt  
integral with the factorisation corresponding to $K^2=s_{ab}\to 0$.}
\label{fig:doublepole}
\end{center}
\end{figure}

A further complication arises for one-loop amplitudes where the external helicities are not identical, the simplest case being the {\it single-minus} 
amplitude with one negative helicity gluon and the rest positive helicity.
Double poles arise in these amplitudes for complex momenta where factorisations as in fig.~\ref{fig:doublepole} occur. 
The factorisation takes the form
\begin{equation}
{V^{(1)}( a^+, b^+, K^+)\over s_{ab}} \times {1 \over s_{ab}} \times A^{(0)}_{n-1:1} (K^-, \cdots ) \sim  { \spb{a}.b \over \spa{a}.b^2 } \times  A^{(0)}_{n-1:1} (K^-, \cdots ) 
\end{equation}
where
\begin{equation}
 {V^{(1)}( a^+, b^+, K^+)\over s_{ab}}=-{i\over 3}{\spb{a}.b\spb{b}.K\spb{K}.a\over s_{ab}}
\end{equation}
is the one-loop three-point vertex~\cite{Bern:2005hs}.
For $n>4$, the all-plus  one-loop amplitude does
not contain double poles since the tree amplitude on the RHS of fig.~\ref{fig:doublepole} vanishes. 
The double poles in the single-minus amplitudes can be seen explicitly in the five-point case~\cite{Bern:1993mq},
\begin{equation}
  A_{5:1}^{(1)} (1^-, 2^+, 3^+, 4^+, 5^+) =
{i \over 3} \, {1\over \spa3.4^2} \biggl[
-{\spb2.5^3 \over
      \spb1.2 \spb5.1} +   {\spa1.4^3 \spb4.5 \spa3.5 \over \spa1.2
      \spa2.3 \spa4.5^2} - {\spa1.3^3 \spb3.2 \spa4.2 \over \spa1.5
      \spa5.4 \spa3.2^2}
      \biggr] 
\end{equation}
where there are $\spa{a}.{b}^{-2}$ singularities for $\spa{a}.b=\spa{2}.3$, $\spa{3}.4$ and $\spa{4}.5$. 

Again the sub-leading in color partial amplitudes can be obtained in terms of the leading in color partial amplitudes using
decoupling identities. 
The naive application of (\ref{eq_COP}) obscures the simplicity of the sub-leading terms. In particular,  there are no double poles in the one-loop
sub-leading partial amplitudes for $n>4$.

To demonstrate this we first consider the 
partial amplitude $A_{n:2}^{(1)}(a_1; b_2, b_3 , b_4 , \cdots b_{n})$.   This can be expressed as a sum over the $A^{(1)}_{n:1}$,
\begin{equation}
A_{n:2}^{(1)}(a_1; b_2, b_3 ,, \cdots b_{n})
=-A_{n:1}^{(1)}(a_1, b_2, b_3  , \cdots b_{n})-A_{n:1}^{(1)}( b_2, a_1, b_3 ,  \cdots b_{n})
\cdots -A_{n:1}^{(1)}(b_2, b_3 , \cdots a_1 , b_{n})
\label{eq:decoX}
\end{equation}
where the sum is over the $n-1$ distinct places where $a_1$ may be inserted within $b_2,b_3\cdots b_n$. 
If we consider the double pole in $\spa{a_1}.{b_2}$ this will only occur in the first two terms and will be of the form
\begin{equation}
-{V^{(1)}( a_1^+, b_2^+, K^+) \over s_{a_1b_2}^2} \times  A^{(0)}_{n-1:1} (K^-,b_3, \cdots b_n )
-{V^{(1)}( b_2^+,a_1^+,  K^+)\over s_{a_1b_2}^2} \times  
A^{(0)}_{n-1:1}  (K^-,b_3, \cdots b_n )\,,
\end{equation}
which vanishes since $V^{(1)}(a^+,b^+,K^+)$ is antisymmetric. 
The double pole in $\spa{b_2}.{b_3}$ also vanishes, but via a different route. Only the second term in (\ref{eq:decoX}) does not contribute and we obtain
\begin{align}
-{V^{(1)}( b_2^+, b_3^+, K^+) \over s_{b_2b_3}^2} \times \biggl( 
A^{(0)}_{n-1:1} (a_1,K^-,b_4, \cdots b_n )+&
A^{(0)}_{n-1:1} (K^-,a_1,b_4, \cdots b_n )
\notag 
\\
&+\cdots 
+A^{(0)} _{n-1:1}(K^-,b_4, \cdots a_1 , b_n )
\biggr) \,.
\label{eq:DP_cancel}
\end{align}
This vanishes due to the decoupling identity for the tree amplitude $A^{(0)} _{n-1:1}$ (\ref{eq:decotree}). 
Similar arguments show the vanishing of double poles for all $A_{n:r}^{(1)}$ with $r > 1$. 

The simplifications in the sub-leading terms allow us to present some compact $n$-point expressions. 
Explicitly, we can find all-$n$ formulae for $A^{(1)}_{n:2}( 1^-; 2^+,\cdots n^+)$  and $A^{(1)}_{n:3}( 1^-,2^+;3^+,\cdots n^+)$:
\begin{eqnarray}
A_{n:2}^{(1)}  (1^-;  2^+,3^+,\cdots ,n^+) =
{ -i \sum_{2 \leq i<j \leq n  } \la 1 | i j |1\ra \over 
\spa2.3\spa3.4 \cdots \spa{(n-1)}.{n} \spa{n}.2 }
\end{eqnarray}
and 
\begin{eqnarray}
A_{n:3}^{(1)}( 1^-, 2^+;  3^+,\cdots ,n^+ ) = \sum_{Z_{(3\cdots n)}}
{ i \sum_{2 \leq i<j \leq n  } \la 1 | i j |1\ra \over 
\spa2.3\spa3.4 \cdots \spa{(n-1)}.{n} \spa{n}.2 }
\end{eqnarray}
where $Z_{(3\cdots n)}$ is the set of cyclic permutations of the set $(3,\cdots n)$.

The vanishing of the $\spa{b_2}.{b_3}$ double poles in (\ref{eq:DP_cancel}) uses a tree level identity, so we do not expect the argument to extend beyond one-loop.  
Specifically if we consider 
$A^{(2)}_{n:2}(a_1; b_2, b_3 , b_4 , \cdots b_{n})$,
a formula for the double pole in $\spa{b_2}.{b_3}$ akin to (\ref{eq:DP_cancel}) will exist but with the tree amplitudes 
$A^{(0)}_{n-1:1}$ replaced by their 
one-loop equivalents $A^{(1)}_{n-1:1}$.  The combination of $A^{(1)}_{n-1:1}$ is that of the decoupling identity (\ref{eq:decoupleA})
so the double pole does not vanish but instead is proportional to
\begin{equation}
{V^{(1)}( b_2^+, b_3^+, K^+) \over s_{b_2b_3}^2} \times  
A^{(1)}_{n-1:2} (a_1; K^-, b_4, \cdots b_n )\,.
\end{equation}

\section{Two-Loop Amplitudes}

A general two-loop amplitude may be expanded in a color trace basis as
\begin{eqnarray}
& & {\cal A}_n^{(2)}(1,2,\cdots ,n) =
N_c^2 \sum_{S_n/Z_n}  \tr(T^{a_1}T^{a_2}\cdots T^{a_n}) A_{n:1}^{(2)}(a_1,a_2,\cdots ,a_n) \notag \\
&+&
N_c\sum_{r=2}^{[n/2]+1}\sum_{S_n/(Z_{r-1}\times Z_{n+1-r}) }   \tr(T^{a_1}T^{a_2}\cdots T^{a_{r-1}})\tr(T^{b_r} \cdots T^{b_n}) 
A_{n:r}^{(2)}(a_1,a_2,\cdots ,a_{r-1} ; b_{r} \cdots b_n)  
\notag \\
&+& \sum_{s=1}^{[n/3]} \sum_{t=s}^{[(n-s)/2]}\sum_{S_n/(Z_s\times Z_t \times Z_{n-s-t})} 
\hskip -1.0 truecm   \tr(T^{a_1}\cdots T^{a_s})\tr(T^{b_{s+1}} \cdots T^{b_{s+t}})
\tr(T^{c_{s+t+1}}\cdots T^{c_n}) 
\notag
\\
& & 
\hskip 7.0truecm 
\times A_{n:s,t}^{(2)}(a_1,\cdots ,a_s;b_{s+1} \cdots b_{s+t} ;c_{s+t+1}\cdots c_n ) 
\notag \\
&+&\sum_{S_n/Z_n}  \tr(T^{a_1}T^{a_2}\cdots T^{a_n}) A_{n:1B }^{(2)}(a_1,a_2,\cdots ,a_n)\,.
\end{eqnarray}
Again, for $n$ even and $r=n/2+1$ there is an extra $Z_2$ in the summation to ensure each color structure only appears once.
In the $s,t$ summations there is an extra $Z_2$ when exactly two of $s$, $t$ and $n-s-t$ are equal and an extra $S_3$ when all three are equal. 

For five-point amplitudes this  reduces to
\begin{eqnarray}
{\cal A}_5^{(2)}(1,2,3,4,5) &=&
N_c^2 \sum_{S_5/Z_5}  \tr(T^{a_1}T^{a_2}T^{a_3}T^{a_4}T^{a_5}) 
A_{5:1}^{(2)}(a_1,a_2,a_3,a_4,a_5 ) 
\notag\\
&+&
N_c\sum_{S_5/Z_4}   \tr(T^{a_1})\tr(T^{b_2}T^{b_3}T^{b_4}T^{b_5}) 
A_{5:2}^{(2)}( a_1 ; b_2,b_3,b_4,b_5 ) 
\notag\\ &+&
N_c \sum_{S_5/(Z_2\times Z_3)} \tr(T^{a_1}T^{a_2})\tr(T^{b_3}T^{b_4}T^{b_5})A_{5:3}^{(2)}(a_1,a_2 ;  b_3,b_4,b_5 )
\notag\\
&+& \sum_{S_5/(Z_2\times Z_3)}   \tr(T^{a_1})\tr(T^{b_2})\tr(T^{c_3}T^{c_4}T^{c_5}) 
A_{5:1,1}^{(2)}( a_1  ; b_2  ;c_3,c_4,c_5 ) 
\notag\\ &+&
\sum_{S_5/( Z_2\times Z_2 \times Z_2)}  \tr(T^{a_1})\tr(T^{b_2}T^{b_3})\tr(T^{c_4}T^{c_5}) A_{5:1,2}^{(2)}( a_1 ;b_2 ,b_3 ;  c_4,c_5 )
\notag\\
&+&\sum_{S_5/Z_5}   \tr(T^{a_1}T^{a_2}T^{a_3}T^{a_4}T^{a_5}) 
A_{5:1B}^{(2)}(a_1,a_2,a_3,a_4,a_5 )
\end{eqnarray}
which for an $SU(N_c)$ gauge group simplifies to 
\begin{eqnarray}
{\cal A}_5^{(2)}(1,2,3,4,5) &=&
N_c^2 \sum_{S_5/Z_5}   \tr(T^{a_1}T^{a_2}T^{a_3}T^{a_4}T^{a_5}) 
A_{5:1}^{(2)}(a_1,a_2,a_3,a_4,a_5 ) 
\notag\\
&+&
N_c \sum_{S_5/(Z_2\times Z_3)} \tr(T^{a_1}T^{a_2})\tr(T^{b_3}T^{b_4}T^{b_5})A_{5:3}^{(2)}(a_1,a_2 ;  b_3,b_4,b_5 )
\notag\\
&+&\sum_{S_5/Z_5}   \tr(T^{a_1}T^{a_2}T^{a_3}T^{a_4}T^{a_5}) 
A_{5:1B}^{(2)}(a_1,a_2,a_3,a_4,a_5 )\,.
\end{eqnarray}
Thus there are three independent functions to be determined: $A_{5:1}^{(2)}$,  $A_{5:3}^{(2)}$ and  $A_{5:1B}^{(2)}$.  By themselves 
the $U(1)$ decoupling identities do not determine any of the three, however they can be used to obtain the specifically $U(N_c)$ functions 
$A_{5:2}^{(2)}$ , $A_{5:1,1}^{(2)}$ and $A_{5:1,2}^{(2)}$: 
\begin{align}
A_{5:2}^{(2)}(1;  2,3,4,5)&=-A_{5:1}^{(2)}( 1,2,3,4,5 )-A_{5:1}^{(2)}( 2,1,3,4,5 )-A_{5:1}^{(2)}( 2,3,1,4,5 )-A_{5:1}^{(2)}( 2,3,4,1,5 )\,,
\notag \\
A_{5:1,1}^{(2)}( 4 ;  5 ; 1,2,3 )
&=
-A_{5:2}^{(2)}( 5 ; 1,2,3,4 )
-A_{5:2}^{(2)}( 5 ; 1,2,4,3 )
-A_{5:2}^{(2)}(  5; 1,4,2,3 )
-A_{5:3}^{(2)}( 4,5 ; 1,2,3 )\notag\\
&=
-A_{5:3}^{(2)}( 4,5 ; 1,2,3 )+\sum_{COP\{4,5\}\{1,2,3\}} A_{5:1}^{(2)}( 1,2,3,4,5 )
\notag\\
\intertext{and}
A_{5:1,2}^{(2)}(1 ; 2,3 ; 4,5) &=
-A_{5:3}^{(2)}(2,3 ; 1,4,5)
-A_{5:3}^{(2)}(2,3 ; 1,5,4)
-A_{5:3}^{(2)}(4,5 ; 1,2,3)
-A_{5:3}^{(2)}(4,5 ; 1,3,2)
\notag\\
&= 0\,. 
 \label{eq:U1decouple}
 \end{align}
Decoupling identities do not relate the $A^{(2)}_{n:1B}$ to the other terms but do impose a tree-like identity, 
\begin{equation}
A_{n:1B}^{(2)}(1,2,3,\cdots n ) +A_{n:1B}^{(2)}(2,1,3,\cdots n )+\cdots A_{n:1B}^{(2)}(2,\cdots, 1, n )=0\,,
\label{eq:decoupleB}\end{equation}
which in itself does not specify $A_{n:1B}^{(2)}$ completely. 
There are however further color restrictions beyond the decoupling identities~\cite{Naculich:2011ep,Edison:2011ta} which may be obtained by recursive methods.  
These, together with eq.~(\ref{eq:decoupleB})  determine the $A_{5:1B}^{(2)}$ in terms of the $A_{5:1}$ and $A_{5:3}$
\begin{eqnarray}
A_{5:1B}^{(2)}(1,2,3,4,5)   &=&  
-A_{5:1}^{(2)}(1, 2, 4, 3,  5) 
+2 A_{5:1}^{(2)}(1, 2, 5, 3, 4) 
+A_{5:1}^{(2)}( 1, 2, 5, 4, 3)
\notag \\
& &-A_{5:1}^{(2)}( 1, 3, 2, 4, 5) 
+2 A_{5:1}^{(2)}( 1, 3, 4, 2, 5) 
-5 A_{5:1}^{(2)}( 1, 3, 5, 2, 4) 
\notag \\
& &-2 A_{5:1}^{(2)}( 1, 3, 5, 4, 2 ) 
+2 A_{5:1}^{(2)}( 1, 4, 2, 3, 5 ) 
+A_{5:1}^{(2)}( 1, 4, 3, 2, 5 ) 
\notag \\
& &+2 A_{5:1}^{(2)}(1, 4, 5, 2, 3)
+A_{5:1}^{(2)}(1, 4, 5, 3, 2 )
\notag \\
&-&\frac{1}{2} \sum_{Z_5(1,2,3,4,5)} \bigg(   A_{5:3}^{(2)}( 1 ,2 ;  3,4,5 ) 
 -A_{5:3}^{(2)}( 1,3 ; 2,4,5 ) \biggr)\,.
\label{eq:Edi}
\end{eqnarray}
Our calculation we determine $A_{5:1B}^{(2)}$ directly and we use (\ref{eq:Edi}) as a consistency check.

\section{Singularity Structure of the All-plus Two-loop Amplitudes}

The IR singular structure of a color partial amplitude is determined by general theorems~\cite{Catani:1998bh}. Consequently we can split the amplitude into
singular terms $U^{(2)}_{n:\lambda}$ and finite terms $F^{(2)}_{n:\lambda}$, 
\begin{eqnarray}
\label{definitionremainder}
A^{(2)}_{n:\lambda} =& U^{(2)}_{n:\lambda}
+  \; F^{(2)}_{n:\lambda}  +   {\mathcal O}(\epsilon)\, .
\end{eqnarray}
As the all-plus
tree amplitude vanishes, $U^{(2)}_{n:\lambda}$ simplifies considerably and is at worst $1/\epsilon^2$.
In general an amplitude has UV divergences, collinear IR divergences and soft IR divergences. 
As the tree amplitude vanishes,  both the UV divergences and collinear IR divergences are proportional to $n$ and cancel leaving only the 
soft IR singular terms~\cite{Kunszt:1994np}. 
The leading case, 
$U^{(2)}_{n:1}$, is proportional to the one-loop amplitude,
\begin{equation}
U_{n:1}^{(2)}  =A_{n:1}^{(1)} \times  I_n^{(2)}
\end{equation}
where
\begin{equation}
I_n^{(2)}= \left[ - \sum_{i=1}^{n} \frac{1}{\epsilon^2} \left(\frac{\mu^2}{-s_{i,i+1}}\right)^{\epsilon}  \right] \,.
\end{equation}
In appendix~\ref{ap:IR} the form of the two-loop IR divergences for the other un-renormalised partial amplitudes are presented
in a color trace basis. 

Given the general expressions for $U_{n:\lambda}^{(2)}$, the challenge is to compute the finite parts of the amplitude: $F_{n:\lambda}^{(2)}$. 
This finite remainder function $F_{n:\lambda}^{(2)}$ can be further split into polylogarithmic and rational pieces,
\begin{equation}
F_{n:\lambda}^{(2)} = \Fcc_{n:\lambda}+R_{n:\lambda}^{(2)}\; .
\end{equation}
We calculate the former piece using four-dimensional unitarity and the latter using recursion.

\section{Unitarity}

\begin{figure}[h]
\centerline{
    \begin{picture}(170,150)(-150,-40) 
    \Text(-24,92)[c]{$a)$}   
     \Line( 0, -24)( 0,60)
     \Text(0,-32)[c]{$_+$}
     \Text(-14,75)[c]{$_+$}
     \Text(74,-15)[c]{$_+$}
     \Text(-32,0)[c]{$_+$}
\Text(-24,-20)[c]{$K_2$}
\Text(84,-20)[c]{$k_1$}
\Text(-24,80)[c]{$k_3$}
\Text(84,80)[c]{$K_4$}
\Vertex(-6,-18){2}
     \Vertex(-18,-6){2}
     \Vertex(-13,-13){2}
     \Line( 0,60)(60,60)
     \Line(60,60)(60, 0)
     \Line(60, 0)( -24, 0)
     \Line( 0,60)(-10,70)
     \Line(60,60)(60,84)
     \Line(60,60)(84,60)
     \Text(60,92)[c]{$_+$}
     \Text(92,60)[c]{$_+$}
     \Vertex(66,78){2}
     \Vertex(78,66){2}
     \Vertex(73,73){2}
     \Line(60, 0)(70,-10)
     \DashLine(52,30)(68,30){2}
     \DashLine(-8,30)(8,30){2}
     \DashLine(30,8)(30,-8){2}
     \DashLine(30,52)(30,68){2}
      \Text(-5,25)[c]{$_-$}
      \Text(-5,35)[c]{$_+$}
      \Text(25,-5)[c]{$_-$}
      \Text(35,-5)[c]{$_+$}
      \Text(65,25)[c]{$_-$}
      \Text(65,35)[c]{$_+$}
      \Text(25,65)[c]{$_-$}
      \Text(35,65)[c]{$_+$}
      \Text(30,-25)[c]{$I_4^{2m\, e}$} 
   \CCirc(60,60){5}{Black}{Purple} 
    \end{picture} 
    \begin{picture}(170,150)(-110,-40)    
     \Text(-24,92)[c]{$b)$} 
     \Line( 0, 0)(30,60)
     \Line(60, 0)(30,60)
     \Line(60, 0)( 0, 0)
     \Line( 0, 0)(-24,0)
     \Line( 0, 0)(0,-24)
     \Text(0,-32)[c]{$_+$}
     \Text(-32,0)[c]{$_+$}
     \Vertex(-6,-18){2}
     \Vertex(-18,-6){2}
     \Vertex(-13,-13){2}
     \Line(30,60)(50,80)
     \Line(30,60)(10,80)
     \Text(55,85)[c]{$_+$}
     \Text(5,85)[c]{$_+$}
     \Text(74,-15)[c]{$_+$}
     \Vertex(40,75){2}
     \Vertex(20,75){2}
     \Vertex(30,78){2}
     \Line(60, 0)(70,-10)
     \DashLine(39,25)(53,33){2}
     \DashLine(21,25)(7,33){2}
     \DashLine(30,8)(30,-8){2}
     \Text(14,37)[c]{$_+$}
     \Text(8,25)[c]{$_-$}
     \Text(46,37)[c]{$_+$}
     \Text(52,25)[c]{$_-$}
     \Text(25,-5)[c]{$_-$}
     \Text(35,-5)[c]{$_+$}
     \Text(30,-25)[c]{$I_3^{2m}$} 
    \CCirc(30,60){5}{Black}{Purple} 
    \end{picture} 
\def\redpicbit{
    \begin{picture}(170,150)(-40,-40)    
     \Text(-24,92)[c]{$c)$} 
     \Line( 0, 0)(30,60)
     \Line(60, 0)(30,60)
     \Line(60, 0)( 0, 0)
     \Line( 0, 0)(-10,-10)
     \Line(30,60)(50,80)
     \Line(30,60)(10,80)
     \Vertex(40,75){2}
     \Vertex(20,75){2}
     \Vertex(30,78){2}
     \Line(60, 0)(70,-10)
     \Text(74,-15)[c]{$_+$}
     \Text(-14,-15)[c]{$_+$}
     \DashLine(39,25)(53,33){2}
     \DashLine(21,25)(7,33){2}
     \DashLine(30,8)(30,-8){2}
     \Text(14,37)[c]{$_+$}
     \Text(8,25)[c]{$_-$}
     \Text(46,37)[c]{$_+$}
     \Text(52,25)[c]{$_-$}
     \Text(25,-5)[c]{$_-$}
     \Text(35,-5)[c]{$_+$}
     \Text(25,5)[c]{$_+$}
     \Text(35,5)[c]{$_-$}
     \Text(30,-25)[c]{$I_3^{1m}$} 
   \CCirc(30,60){5}{Black}{Purple} 
    \end{picture} 
}
\redpicbit
    \begin{picture}(170,150)( 30,-40)  
     \Text(-24,92)[c]{$d)$} 
     \CArc(30,30)(20,0,360)
     \Line(10,30)( -12,45)
     \Line(10,30)( -12,15)
     \Text(-15,48)[c]{$_+$}
     \Text(-15,12)[c]{$_+$}
     \Vertex(-8,38){2}
     \Vertex(-10,30){2}
     \Vertex(-8,22){2}
     \Line(50,30)(72,45)
     \Line(50,30)(72,15)
     \Text(75,48)[c]{$_+$}
     \Text(75,12)[c]{$_+$}
     \Vertex(68,38){2}
     \Vertex(70,30){2}
     \Vertex(68,22){2}
    \CCirc(50,30){5}{Black}{Purple}
     \DashLine(30,43)(30,59){2}
     \DashLine(30,17)(30,1){2}
     \Text(25,55)[c]{$_-$}
     \Text(25,5)[c]{$_-$}
     \Text(35,55)[c]{$_+$}
     \Text(35,5)[c]{$_+$}   
     \Text(30,-25)[c]{$I_2$} 
 \end{picture} 
    }
    \caption{Four dimensional cuts of the two-loop all-plus amplitude involving an all-plus one-loop vertex 
    (indicated by $\bullet\;$ ). In the boxes $K_2$ may be null but $K_4$ must contain at least two external legs. } 
    \label{fig:oneloopstyle}
\end{figure}
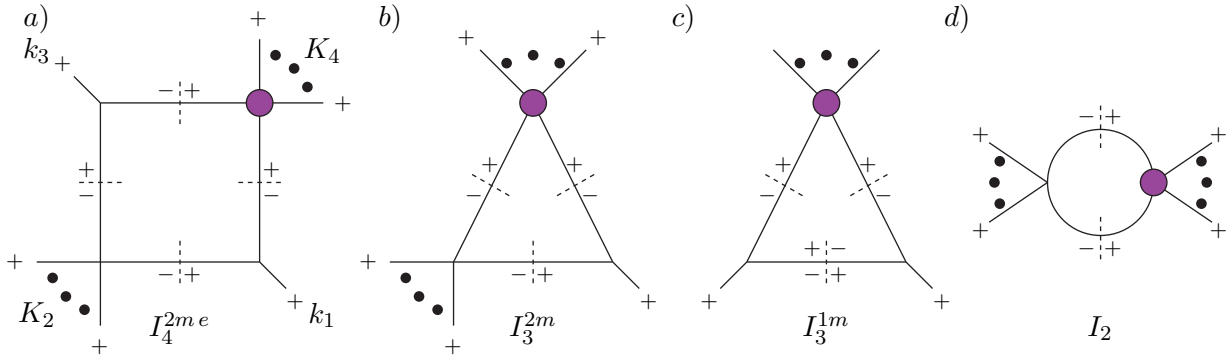

$D$-dimensional unitarity techniques can be used to generate the integrands~\cite{Badger:2013gxa} for the five-point amplitude which can then be 
integrated to obtain the amplitude~\cite{Gehrmann:2015bfy}.   However the organisation of the amplitude in the previous section allows us to obtain the finite 
polylogarithms using four-dimensional unitarity~\cite{Bern:1994zx,Bern:1994cg} where the cuts are evaluated in four dimension with the corresponding simplifications.  
With this simplification the all-plus one-loop amplitude effectively becomes an additional on-shell vertex and the two-loop cuts effectively become one-loop 
cuts with a single insertion of this vertex.  The non-vanishing four dimensional cuts are shown in fig.~\ref{fig:oneloopstyle}. 

The cuts allow us to determine the coefficients, $\alpha_i$, of box, triangle and bubble functions in the amplitude. The integral functions are
\begin{equation}
I_2(K^2)={(-K^2)^{-\epsilon}\over \epsilon(1-2\epsilon)}\,,
\end{equation}

\begin{equation}
I_{3}^{1\rm m}(K^2) = {1\over\epsilon^2} (-K^2)^{-1-\epsilon}\,,
\;\;\;\; 
I_{3}^{2 \rm m}\bigl(K_1^2,K_2^2\bigr)= {1\over\epsilon^2}
{(-K_1^2)^{-\epsilon}-(-K_2^2)^{-\epsilon} \over  (-K_1^2)-(-K_2^2) }\, ,
\end{equation}
and
\begin{eqnarray}
I^{\rm 2m \hskip 1pt e}_4(S,T,K^2_2,K_4^2) & = &-{2  \over S T-K_2^2K_4^2 }
\Biggl[-{1\over\epsilon^2} \Bigl[ (-S)^{-\epsilon} +
(-T)^{-\epsilon} - (-K_2^2)^{-\epsilon} - (-K_4^2)^{-\epsilon}\Bigr] \cr
\notag \\ 
  + \Li_2\left(1-{ K_2^2 \over S }\right)
 &  +& \ \Li_2\left(1-{K_2^2 \over T}\right)
 + \Li_2\left(1-{ K_4^2 \over S }\right)
   + \ \Li_2\left(1-{K_4^2 \over T}\right)  
\notag \\ 
& &  -\Li_2\left(1-{K_2^2K_4^2\over S T}\right)
   +{1\over 2} \ln^2\left({ S  \over T}\right)
\Biggr] 
\end{eqnarray}
where $S=(k_1+K_2)^2$ and $T=(k_1+K_4)^2$. 

\def\plF{{\rm F}}

The bubbles in principle would determine the $(-s)^{\epsilon}/\epsilon$
infinities.   However,  explicit calculation using, for example,  a canonical basis approach~\cite{Dunbar:2009ax}  shows that they have zero coefficient.
This is a property of this particular helicity configuration and is due to the vanishing of the tree amplitude. 
The triangles only  contribute to 
$U_{n:\lambda}^{(2)}$, while the box functions contribute 
to both $U_{n:\lambda}^{(2)}$ and the finite polylogarithms. 
Separating these pieces we have 
\begin{equation}
I_4^{2m\hskip 1pt e}(S,T,K^2_2,K_4^2)  = I_4^{2m\hskip 1pt e}\bigg|_{IR}  \;\; -{2 \over S T-K_2^2K_4^2 } \plF^{2m}[ S,T,K^2_2,K_4^2] 
\end{equation}
where $F^{2m}$ is a dimensionless combination of polylogarithms.

The IR terms combine to give the correct IR singularities~\cite{Dunbar:2016cxp}, 
\begin{eqnarray}&
\biggl(   
\sum  \alpha_{i} I_{4,i}^{\rm 2m\hskip 1pt e } 
\biggr\vert_{IR}
+\sum  \alpha_{i}  I_{3,i}^{2 \rm m}  
+\sum  \alpha_{i}  I_{3,i}^{1 \rm m} 
\biggr)_{\lambda} 
= 
U^{(1),\epsilon^0}_{n:\lambda}(1^+,2^+,\cdots, n^+)
\label{eq:UniCheck}
\end{eqnarray}
where $U^{(2),\epsilon^0}_n(1^+,2^+,\cdots, n^+)$ is the order $\epsilon^0$ truncation.   
    We have checked the relation of (\ref{eq:UniCheck}) by using four dimensional unitarity techniques to 
compute the coefficients and then comparing to the expected form of $U^{(2)}_n$ given in appendix~\ref{ap:IR} for $n$ up to 9 points.

The remaining parts of the box integral functions generate the finite polylogarithms.
The expression for $\Fcc_n$ is~\cite{Dunbar:2016cxp} 
\begin{equation}
P_n^{(2)}   =  \sum_{i}   c_{i}  \plF^{2m}_{i}
\end{equation}
where
\begin{eqnarray}
\plF^{\rm 2m }[S,T,K^2_2,K_4^2]  &=& 
  \Li_2\left(1-{ K_2^2 \over S }\right)
   +\ \Li_2\left(1-{K_2^2 \over T}\right)
 + \Li_2\left(1-{ K_4^2 \over S }\right)
   + \ \Li_2\left(1-{K_4^2 \over T}\right)  
\notag \\ 
& &  -\Li_2\left(1-{K_2^2K_4^2\over S T}\right)
   +{1\over 2} \ln^2\left({ S  \over T}\right)
\end{eqnarray}
and, in the specific case where $K_2^2=0$, 
\begin{eqnarray}
\plF^{\rm 1m }[S,T,K_4^2] &\equiv & \plF^{\rm 2m }[S,T,0,K_4^2]
\notag\\  &=& 
 \Li_2\left(1-{ K_4^2 \over S }\right)
   + \ \Li_2\left(1-{K_4^2 \over T}\right)  
   +{1\over 2} \ln^2\left({ S  \over T}\right)+{\pi^2 \over 6} \; .
\end{eqnarray}

Let us now consider the specific five-point case where only the $K^2_2=0$ case occurs.  
\begin{figure}[H]
\centerline{
    \begin{picture}(170,120)(-10,-20)    
     \ArrowLine( 0, 0)( 0,60)
     \ArrowLine(60,60)( 0,60)
     \ArrowLine(60,60)(60, 0)
     \ArrowLine( 0, 0)(60, 0)
     \Line( 0, 0)(-8,-8)
     \Line( 0,60)(-8,68)
     \Line(60,60)(60,75)
     \Line(60,60)(75,60)
     \Line(60, 0)(68,-8)
     \Text(-12,-12)[c]{$b^+$}  
     \Text(-12, 72)[c]{$c^+$}  
     \Text( 72,-12)[c]{$a^+$}  
     \Text( 85, 62)[c]{$e^+$}  
     \Text( 60, 82)[c]{$d^+$} 
     \Text( 60, 60)[c]{$\bullet$}
     \CCirc(60,60){5}{Black}{Purple}
     \Text(-15, 30)[c]{$\ell_3,p$}  
     \Text( 75, 30)[c]{$\ell_1,m$}  
     \Text( 30,-10)[c]{$\ell_2,n$}  
     \Text( 30, 70)[c]{$\ell_4,q$}
     \Text( 50, 65)[c]{$^+$}
     \Text( 10, 65)[c]{$^-$}     
     \Text( 50, -5)[c]{$^+$}
     \Text( 10, -5)[c]{$^-$}     
     \Text( 65, 50)[c]{$^+$}     
     \Text( 65, 10)[c]{$^-$}     
     \Text( -5, 50)[c]{$^+$}     
     \Text( -5, 10)[c]{$^-$}     
    \end{picture} 
    }
    \caption{The labelling and internal helicities of the quadruple cut.}
    \label{fig:oneloopbox}
\end{figure}
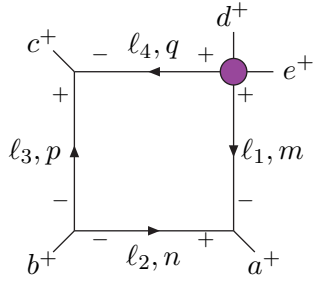
In this case the one-loop corner is a four-point amplitude and the color partial amplitudes simplify since
\begin{equation}
A_{4:1}^{(1)}(1,2,3,4)=A_{4:1}^{(1)}(1,2,4,3)=A_{4:1}^{(1)}(1,3,2,4)\,,
\end{equation}
which implies that
\begin{equation}
A_{4:2}^{(1)}(1;2,3,4)=-3 A_{4:1}^{(1)}(1,2,3,4)
\;\;{\rm and}\;\;
A_{4:3}^{(1)}(1,2;3,4)=6A_{4:1}^{(1)}(1,2,3,4)
\end{equation}
so that the full color amplitude factorises into color and kinematic terms
\begin{equation}
{\cal A}^{(1)}_4 (\ell_1,\ell_4, d, e )
={\cal C} \times A_{4:1}^{(1)}(\ell_4,d,e,\ell_1)\,.
\end{equation}

Since the three-point tree amplitudes also factorise, the quadruple cut of this box function factorises as
\begin{equation}
{\cal C'} \times 
A_{3}^{(0)} (a,\ell_2,\ell_1)  
A_{3}^{(0)} (b,\ell_3,\ell_2)  
A_{3}^{(0)} (c,\ell_4,\ell_3)  
A_{4:1}^{(1)}(\ell_4,d,e,\ell_1)\,.
\end{equation}

The solution to the quadruple cuts in this case is
\begin{eqnarray}
\ell_2 &=& -\frac{\spa{b}.a}{\spa{c}.a} \bar\lambda_b \lambda_c
, \;\;\;\;\;\;\;\;\;\;\;\;\;\;
\ell_3 = -\frac{\spa{b}.c}{\spa{a}.c} \bar\lambda_b \lambda_a\,,
\notag \\
\ell_1 &=&{ \spa{a}.c \bar\lambda_a+\spa{b}.c \bar \lambda_b \over \spa{a}.c }  \lambda_a 
, \; \; 
\ell_4 ={ \spa{c}.a \bar\lambda_c+\spa{b}.a \bar \lambda_b \over \spa{c}.a }  \lambda_c \,.
\end{eqnarray}
So that 
\begin{align}
A_{3}^{(0)} (a,\ell_2,\ell_1)  
A_{3}^{(0)} (b,\ell_3,\ell_2)  
A_{3}^{(0)} (c,\ell_4,\ell_3)  
A_{4:1}^{(1)}(\ell_4,d,e,\ell_1) &=  {2i\over 3} s_{ab}s_{bc} \times  { \spb{d}.e^2 \over \spa{a}.b \spa{b}.c \spa{c}.a } 
\notag 
\\
&=  {1\over 3} s_{ab}s_{bc} \times A_{5:3}^{(1)}(d,e; a, b, c)\,.
\end{align}
\def\Nc{N_c}
Consequently,
\begin{equation}
P^{(2)}_{5:\lambda} \sim \sum   A_{5:3}^{(1)}(d,e; a,b,c) \times \plF^{1m}_{abc;de} 
\end{equation}
where $\plF^{1m}_{abc;de}\equiv \plF^{1m}[s_{ab},s_{bc},s_{de}]$.
We can determine the terms in the summation by expanding ${\cal C}'$ using $U(N_c)$ identities:
\begin{align}
{\cal C}'_{(de; abc)} =& \sum_{m,n,p,q}
\Bigl( (\tr[amn]-\tr[man] )(\tr[bpn]-\tr[pbn])(\tr[pcq]-\tr[pqc]) \Bigr)
\notag \\
 \times & 
\Bigl( N_c\tr[mqed]+N_c\tr[meqd]/2+N_c\tr[qemd]/2+N_c\tr[qmed]
\notag \\
 & 
-3\tr[m]\tr[qde]-3\tr[q]\tr[mde]
-3\tr[d]\tr[emq]-3\tr[d]\tr[eqm]
\notag \\
 & 
+3\tr[de]\tr[mq]+3\tr[dm]\tr[eq] +3\tr[dq]\tr[em] 
+\{ d \leftrightarrow e \}
\Bigr)
\notag \\
=\Nc^2  \Big(&\tr[deabc]+\tr[edabc]-\tr[badec]-\tr[baedc]\Big) 
\notag \\
  +\Nc \Big(&-2\tr[d](\tr[eabc]-\tr[baec])-2\tr[e](\tr[dabc]-\tr[badc])
  \notag \\
     &-\tr[a](\tr[debc]+\tr[edbc]-\tr[bdec]-\tr[bedc])\notag \\
     &-\tr[b] (\tr[deac]+\tr[edac]-\tr[aedc]-\tr[adec])\notag \\
     &-\tr[c] (\tr[deab]+\tr[edab]-\tr[adeb]-\tr[aedb])\notag \\
 + 8 &\tr[de] (\tr[abc]-\tr[bac])+\tr[da](\tr[bec]-\tr[ebc])\notag \\
  +&\tr[db](\tr[aec]-\tr[eac])+\tr[dc](\tr[aeb]-\tr[eab])\notag \\
   - & \tr[ea](\tr[dbc]-\tr[bdc])-\tr[eb](\tr[dac]-\tr[adc]) -\tr[ec](\tr[dab]-\tr[adb])\Big)\notag \\
     +3 \Big(-2&\tr[d]\tr[e](\tr[abc]-\tr[bac])+\tr[d]\tr[a](\tr[ebc]-\tr[bec])\notag \\
     +&\tr[d]\tr[b](\tr[eac]-\tr[aec])+\tr[d]\tr[c](\tr[eab]-\tr[aeb])\notag \\
     +&\tr[e]\tr[a](\tr[dbc]-\tr[bdc])+\tr[e]\tr[b](\tr[dac]-\tr[adc])\notag \\
     +&\tr[e]\tr[c](\tr[dab]-\tr[adb])\Big)
     \notag \\
     +6\Big(
     \tr[deabc]&-\tr[dcbae]
     +\tr[dcbea]-\tr[daebc]
     +\tr[dceba]-\tr[dabec]
     +\tr[dcaeb]-\tr[dbeac]\notag \\
     +\tr[dbaec]&-\tr[dceab]
     +\tr[dabce]-\tr[decba]
     +\tr[daecb]-\tr[dbcea]
     +\tr[dbeca]-\tr[daceb]\Big)\,.
\end{align}

This is an expansion of the form 
\begin{equation}
C'_{(de; abc)}= \sum_{\lambda} a_{(de; abc)}^{\lambda}  C^{\lambda}
\end{equation}
where the $C^\lambda$ are the different color structures.  Consequently the polylogarithmic part of the partial amplitudes is
\begin{equation}
P^{(2)}_{5:\lambda} =   \sum_{(de; abc)} a_{(de; abc)}^{\lambda}  A_{5:3}(d,e; a,b,c) \times \plF^{1m}_{abc;de}\,.
\end{equation}
Specifically we recover the previous results of~\cite{Gehrmann:2015bfy} and \cite{Badger:2019djh}.
Defining $S_{5:1}=Z_5(a,b,c,d,e)$, $S_{5:2}=Z_{4}(b,c,d,e)$ and $S_{5:3}=Z_2(a,b)\times Z_3(c,d,e)$ we have

\begin{eqnarray}
P^{(2)}_{5:1}(a,b,c,d,e) &=& \sum_{S_{5:1}} - A^{(1)}_{5:3}(d,e; a,b,c) \plF^{1m}_{abc;de}\,,
\notag \\
P^{(2)}_{5:3}(a,b:c,d,e) &=&    \frac43\sum_{S_{5:3}} \Bigl(A^{(1)}_{5:3}(a,b ; c,d,e)\;\plF^{1m}_{cde,ab}\notag\\
  &+ &  \frac14 A^{(1)}_{5:3}(a,c;b,e,d)\;(\plF^{1m}_{bed;ac}+\plF^{1m}_{bde;ac}-\plF^{1m}_{dbe;ac})\Bigr)\,.
\label{eq:subleadin}
\end{eqnarray}
We also determine directly the remaining $SU(N_c)$  partial amplitude, 
\begin{align}
    P^{(2)}_{5:1B}(a,b,c,d,e)= 2\sum_{S_{5:1}} \Bigl(&A^{(1)}_{5:3}(a,b ; c,d,e)\;\plF^{1m}_{cde;ab}+\notag\\
   & A^{(1)}_{5:3}(a,c;b,e,d)\;(\plF^{1m}_{bed;ac}+\plF^{1m}_{bde;ac} -\plF^{1m}_{dbe;ac} )\Bigr)\,.
\end{align}
This expression matches that obtained by using the results of (\ref{eq:subleadin}) in (\ref{eq:Edi}). 

The specifically $U(N_c)$ partial amplitudes may also be extracted directly:
\begin{align}
    P^{(2)}_{5:2}(a;b,c,d,e)= -\frac{2}{3}\sum_{S_{5:2}} 
    \Big(&A^{(1)}_{5:3}(a,b;c,d,e)\;\plF^{1m}_{cde; ab} \notag\\
    &+\frac{1}{2}A^{(1)}_{5:3}(b,c;a,e,d)\;(\plF^{1m}_{ade; bc}+\plF^{1m}_{dea;bc}-\plF^{1m}_{dae ;bc})\Big)
\end{align}
and
\begin{align}
    P^{(2)}_{5:1,1}(a;b;c,d,e)= -\sum_{S_{5:3}} \Bigl(&A^{(1)}_{5:3}(a,b; c,d,e)\;\plF^{1m}_{cde ;ab}\notag\\
   & +A^{(1)}_{5:3}(a,c;b,e,d)\;(
   \plF^{1m}_{bed; ac}+
   \plF^{1m}_{bde; ac}-
   \plF^{1m}_{dbe; ac} )\Bigr)\,.
\end{align}

As a check we have confirmed that these satisfy the decoupling identities (\ref{eq:U1decouple}).

\section{Recursion}

The remaining part of the amplitude is the rational function $R_{n:\lambda}^{(2)}$.  
In \cite{Dunbar:2017nfy} we described a technique for evaluating this for the leading in color partial amplitude. 
We review this here and describe the extensions necessary to determine the full-color amplitude. 

As $R_{n:\lambda}^{(2)}$ is a rational function we can obtain it recursively
given sufficient information about its singularities.
Britto-Cachazo-Feng-Witten (BCFW) \cite{Britto:2005fq} exploited the analytic properties of $n$-point tree amplitudes under
a complex shift of their external momenta to compute these amplitudes recursively.
Explicitly the  BCFW shift acting on two momenta, say  $p_\ki$ and $p_\kj$, is
\begin{equation}
\bar\lambda_{{\ki}}\to \bar\lambda_{\hat{\ki}} =\bar\lambda_\ki - z \bar\lambda_\kj 
,\;\;
\lambda_{{\kj}}\to\lambda_{\hat{\kj}} =\lambda_\kj + z \lambda_\ki \; .
\end{equation}
This introduces a complex parameter, $z$, whilst  preserving  overall momentum conservation and keeping all external momenta null.  
Alternative shifts can also be employed, for example 
the Risager shift~\cite{Risager:2005vk} which acts on three momenta, say $p_a$, $p_b$ and $p_c$, to give
\begin{equation}
    \begin{split}
         \lambda_a\to\lambda_{\hat{a}}=\lambda_a+z\spb{b}.c\lambda_\eta\,,\\
         \lambda_b\to\lambda_{\hat{b}}=\lambda_b+z\spb{c}.a\lambda_\eta\,,\\
         \lambda_c\to\lambda_{\hat{c}}=\lambda_c+z\spb{a}.b\lambda_\eta\,,
   \end{split}
   \label{Kasper}
\end{equation}
where $\lambda_\eta$ must satisfy $\spa{a}.{\eta}\neq 0$ etc., but is otherwise unconstrained.

After applying either of these shifts, the rational quantity of interest is a complex function parametrized by $z$ i.e. $R(z)$. 
If $R(z)$ vanishes at large $\vert z\vert$, then
Cauchy's theorem applied to $R(z)/z$ over a contour at infinity implies
\begin{equation}
R= R(0)= -\sum_{z_j\neq 0} {\rm Res}\Bigl[ {R(z)\over z}\Bigr]\Bigr\vert_{z_j}\;.
 \label{AmpAsResidues2}
\end{equation}
If the function only contains simple poles, ${\rm Res}\bigl[ {R(z)/z}\bigr]\bigr\vert_{z_j} = {\rm Res}[ R(z)]\bigr\vert_{z_j}/z_j$
and we can use factorisation theorems to determine the residues.
Higher order poles do not present a problem mathematically, for example, given  a function with a 
double pole at $z=z_j$  and its 
Laurent expansion,
\begin{eqnarray}
R(z) &=& \frac{c_{-2}}{(z-z_{j})^2}+ \frac{c_{-1}}{(z-z_{j})}+\mathcal{O}((z-z_{j})^0)\; , \notag \\
\end{eqnarray}
the residue is simply
\begin{equation}
{\rm Res}\Bigl[ {R(z)\over z}\Bigr]\Bigr\vert_{z_j}  = 
-\;\frac{c_{-2}}{z_{j}^2}+ \frac{c_{-1}}{z_{j}}\,.
\end{equation}
To determine this we need to know both the leading and sub-leading poles. As discussed above, loop amplitudes can contain double poles and,
at this point, there are no general theorems determining the sub-leading pole. 

Both the BCFW and Risager shifts break cyclic symmetry of the amplitude by acting on specific legs and the Risager shift further 
introduces the arbitrary spinor $\eta$.
While it is hard to determine {\it a priori} the large $z$  behaviour of an unknown amplitude, recovering cyclic symmetry (and $\eta$ independence)
are powerful checks. 
For the two-loop all-plus amplitude this symmetry recovery does not occur for the BCFW shift (the one-loop all-plus amplitudes
have the same feature). However, symmetry is recovered if we employ the Risager shift~(\ref{Kasper}).

The Risager shift excites poles corresponding to tree:two-loop  and one-loop:one-loop factorisations. The former involve only single poles and their contributions
are readily obtained from the rational parts of the four-point two-loop amplitude~\cite{Bern:2002tk}:

\begin{align*}
    R_{4:1}^{(2)}(K^+,c^+,d^+,e^+)&=\frac13A^{(1)}_{4:1}(K^+,c^+,d^+,e^+)\left(\frac{s_{ce}^2}{s_{cd}s_{de}}+8\right)\,,\\
    R_{4:3}^{(2)}(K^+,c^+;d^+,e^+)&=\frac{1}{9}A^{(1)}_{4:3}(K^+,c^+;d^+,e^+)\left(\frac{s_{cd}^2}{s_{ce}s_{de}}+\frac{s_{ce}^2}{s_{cd}s_{de}}+\frac{s_{de}^2}{s_{cd}s_{ce}}+24\right).
\end{align*}

\begin{figure}[h]
\begin{center}
\includegraphics[scale=0.8]{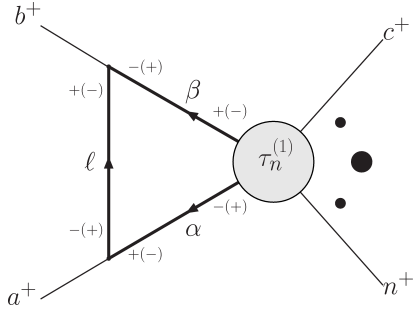}
\end{center}
\caption{Diagram containing the leading and sub-leading poles as $s_{ab}\to 0$. The axial gauge construction permits the off-shell continuation of the internal 
legs.}
\label{fig:axialg}
\end{figure}

The one-loop:one-loop factorisations involve double poles and we need to determine the sub-leading pieces.
By considering a diagram of the form fig.~\ref{fig:axialg} using an axial gauge formalism, we can determine the full pole structure of the rational piece, 
including the non-factorising simple poles.
We have used this approach previously to compute one-loop~\cite{Dunbar:2010xk,Alston:2015gea,Dunbar:2016dgg}
and two-loop amplitudes~\cite{Dunbar:2016aux,Dunbar:2016gjb,Dunbar:2017nfy},
we labelled this process  {\em augmented recursion}. 
In axial gauge formalism  helicity labels can be assigned to internal off-shell legs 
and vertices expressed in terms of nullified momenta~\cite{Schwinn:2005pi,Kosower:1989xy},
\begin{align}
    K^{\flat}=K-\frac{K^2}{2K.q}q
\end{align}
where $q$ is a reference momentum. 
The two off-shell legs  are,
\begin{equation}
    \alpha=\alpha(\ell)=\ell+a\;{\rm and}\; \beta=\beta(l)=b-\ell\,,
\end{equation}
where we also define the sum of these legs, $P_{\alpha \beta}=\alpha+\beta=a+b=P_{ab}$, which is independent of the loop momentum, $\ell$.

The principal helicity assignment in fig.~\ref{fig:axialg}, gives
\begin{eqnarray}
\int\!\! d {\Lambda}^{c}(\alpha^+,a^+,b^+,\beta^-)\;\tau_{n}^{(1),c} (\alpha^{-},\beta^{+},c^+,...,n^+)
\label{eq:tauintinitA}
\end{eqnarray}
where 
\begin{equation}
\int\!\! d\Lambda^c(\alpha^+,a^+,b^+,\beta^-)  \equiv \frac{i}{c_\Gamma(2\pi)^D}\int\!\! \frac{d^D\ell}{\ell^2\alpha^2\beta^2} 
{\cal V}_3(\alpha,a,\ell) {\cal V}_3 ( \ell,b,\beta) \,,
\end{equation}
the vertices are in axial gauge and $\tau_{n}^{(1),c}$ is a doubly off-shell current.

As we are only interested in the residue on the $s_{ab}\to 0$ pole, we do not need the exact current.
It is sufficient that the approximate current satisfies two conditions~\cite{Alston:2015gea,Dunbar:2016aux}:
\begin{enumerate}
    \item[(C1)] The current contains the leading singularity as $s_{\alpha\beta}\to0$ with $\alpha^2,\beta^2\neq0$,
    \item[(C2)] The current is the one-loop, single-minus amplitude in the on-shell limit $\alpha^2,\beta^2\to0$, $s_{\alpha\beta}\neq0$.
\end{enumerate} 
This process is detailed in~\cite{Dunbar:2017nfy}.

We now apply the method to the full color amplitude. 
The  $U(N_c)$ color decomposition of $d\Lambda^c$ contains a common kinematic factor so we have the color decompositions
\begin{equation}
\tau_n^{(1),c}  = \sum _{\lambda} C_{\lambda} \tau^{(1)}_{n:\lambda}  
\;\;\; {\rm and} \;\; 
\int d \Lambda^c =  C_{\Lambda} \int d \Lambda_0 
\end{equation}
where 
\begin{equation}
\int d \Lambda_0(\alpha^+,a^+,b^+,\beta^-) =\frac{i}{c_\Gamma(2\pi)^D}\int\!\! \frac{d^D\ell}{\ell^2\alpha^2\beta^2} 
\frac{[a|\ell|q\ra[b|\ell|q\ra }{\spa a.q\spa b.q} \frac{\spa \beta.q^2}{\spa\alpha.q^2}\,.
\end{equation}
Hence the full color contribution is
\begin{eqnarray}
\sum_{\lambda} C_{\Lambda} C_{\lambda} \int\!\! d \Lambda_0 (\alpha^+,a^+,b^+,\beta^-)\;
\tau_{n:\lambda}^{(1)} (\alpha^{-},\beta^{+},c^+,\cdots n^+).
\label{eq:fullcolortau}
\end{eqnarray}

The various $\tau^{(1)}_{n:\lambda}$ 
can be expressed as  sums of the leading amplitudes $\tau^{(1)}_{n:1}$ via a series of $U(1)$ decoupling identities.

We now focus on the five-point case,  where there are  two distinct forms of the leading current, 
\begin{align}
    \tau^{(1)}_{5:1}(\alpha^-,\beta^+,c^+,d^+,e^+) \;\; {\rm and} \;\; \tau^{(1)}_{5:1}(\alpha^-,c^+,\beta^+,d^+,e^+)\,,
\end{align}
which we call the 'adjacent' and 'non-adjacent' leading currents respectively.

 $\tau^{(1)}_{5:1}(\alpha^-,\beta^+,c^+,d^+,e^+)$ has been calculated 
previously for a specific choice of the axial gauge spinor $\lambda_q=\lambda_d$~\cite{Dunbar:2016aux}.  
Since we require currents for which all the legs have been permuted it is necessary to derive this current for 
arbitrary $\lambda_q$.  The non-adjacent case has not previously been considered. 
The derivation of the adjacent current is given in Appendix B. This current is given by
\begin{equation}
    \begin{split}
       &\tau^{(1)}_{5:1}(\alpha^-,\beta^+,c^+,d^+,e^+) =
       \mathcal{F}^{\;cde}_{dp}\left[1+s_{\alpha\beta}\left(\frac{\spb{q}.e}{\spb{c}.e[q|P_{\alpha\beta}|c\ra}+\frac{[c|q|d\ra}{[q|P_{\alpha\beta}|q\ra[c|e|d\ra}+\frac{[e|q|d\ra}{[q|P_{\alpha\beta}|q\ra[e|c|d\ra}\right)\right]\\
      &+\frac{i}{3\spa{c}.d^2}\frac{\spa{\alpha}.q^2}{\spa{\beta}.q^2}\left[\frac{\spa{\alpha}.c[c|\beta|d\ra}{\spa{d}.e\spa{e}.\alpha}
       +\frac{\spa{c}.e\spb{d}.e}{\spa{d}.e^2}\left(\frac{[q|P_{\alpha\beta}|d\ra^3}{[q|P_{\alpha\beta}|q\ra^3}
       \frac{\spa{q}.c\spa{q}.\alpha[q|\alpha|q\ra}{\spa{\alpha}.c[q|P_{\alpha\beta}|c\ra}
       -3\frac{\spa{q}.d[q|P_{\alpha\beta}|d\ra^2[q|\beta|q\ra}{[q|P_{\alpha\beta}|q\ra^2[q|P_{\alpha\beta}|c\ra}\right)\right]\\
       &+\mathcal{F}^{\;cde}_{sb}+\frac{i}{3\spa{c}.d^2}\left(-\frac{\spb{\beta}.e^2\spb{q}.e}{\spb{e}.{\alpha}\spb{\alpha}.q}
       +[e|q|\alpha\ra\frac{(\spb{e}.{\beta}\spb{\beta}.q[q|P_{\alpha \beta}|q\ra-\spb{\beta}.q^2[e|P_{\alpha\beta}|q\ra)}
       {\spb{\alpha}.q[q|P_{\alpha\beta}|q\ra^2}\right) \\
       &\hskip9.0truecm+{\cal O}(\spa{\alpha}.\beta)+\mathcal{O}(\alpha^2)+\mathcal{O}(\beta^2)
    \end{split}
    \label{eq:adjacentleading}
\end{equation}
where
\begin{align}
    \mathcal{F}^{\;cde}_{dp}       &=\frac{i}3\frac{\spa{\alpha}.q^2}{\spa{\beta}.q^2}\frac{\la q|\alpha\beta|q\ra}{s_{\alpha\beta}}
    \frac{\spa{e}.c\spb{c}.e^3}{\spa{c}.d\spa{d}.e[e|P_{\alpha\beta}|q\ra[c|P_{\alpha\beta}|q\ra}
\\
\intertext{ and}
 \mathcal{F}^{\;cde}_{sb}&=-\frac{i}{3}\frac{[e|P_{\alpha\beta}|\alpha\ra \spb{\beta}.q^2}{\spb{\alpha}.q[q|P_{\alpha\beta}|q\ra}
 \frac{1}{s_{\alpha\beta}}\frac{[e|P_{\alpha\beta}|q\ra}{\spa{c}.d^2}\,.
\end{align}
Setting $\lambda_q=\lambda_d$ in \eqref{eq:adjacentleading} reproduces the current presented in \cite{Dunbar:2016aux}.

The non-adjacent leading current is
\begin{equation}
   \tau^{(1)}_{5:1}(\alpha^-,c^+,\beta^+,d^+,e^+)=\frac{i}{3}\frac{\spa{\alpha}.q^2}{\spa{\beta}.q^2}\left(\frac{\spa{\alpha}.e\spb{e}.c}{\spa{c}.\alpha\spa{d}.e^2}-\frac{\spb{e}.c^3}{[e|\alpha|d\ra[c|\alpha|d\ra} \right)+\mathcal{O}(\spa{\alpha}.\beta).
   \label{eq:nonadjc}
\end{equation}

These currents must be integrated before extracting the rational pole. 
The non-adjacent case integrates to the simple form,
\begin{equation}
    \int\!\!\frac{d^D\ell}{\ell^2\alpha^2\beta^2} 
       \frac{i}3 \frac{[a|\ell|q\ra[b|\ell|q\ra }{\spa{a}.q\spa{b}.q}\frac{\spa{a}.e\spb{e}.c}{\spa{c}.a\spa{d}.e^2}=\frac{i}6\frac{\spb{e}.c\spa{a}.e\spb{a}.b}{\spa{d}.e^2\spa{c}.a\spa{a}.b}.
       \label{eq:nonadjcurrent}
\end{equation}
where the second term in eq.~(\ref{eq:nonadjc}) has been dropped since it is a quadratic pentagon and does not contain any rational terms.
The integrated adjacent case is a generalisation of the previous result~\cite{Dunbar:2016aux}.
Summing over all the channels excited by the Risager shift we recover the full two-loop color decomposition. We present compact forms of 
the $SU(N_c)$ rational pieces below, including the first compact 
form for the rational piece of the maximally non-planar amplitude obtained via a direct computation. 
We find complete agreement with previous calculations \cite{Badger:2019djh} and $R^{(2)}_{5:1B}$ satisfies the constraint \eqref{eq:Edi}.

\begin{equation}
    R^{(2)}_{5:1}(a^+,b^+,c^+,d^+,e^+)=\frac{i}{9}\frac1{\spa{a}.b\spa{b}.c\spa{c}.d\spa{d}.e\spa{e}.a}
    \sum_{S_{5:1}}\Bigl(\frac{{\rm tr}^2_+[deab]}{s_{de}s_{ab}}+5s_{ab}s_{bc}+s_{ab}s_{cd}\Bigr)\,,
\end{equation}
\begin{equation}
    R^{(2)}_{5:3}(a^+,b^+;c^+,d^+,e^+)=\frac{2i}3\frac1{\spa{a}.b\spa{b}.a\spa{c}.d\spa{d}.e\spa{e}.c}
    \sum_{S_{5:3}}\left(\frac{{\rm tr}_-[acde]{\rm tr}_-[ecba]}{s_{ae}s_{cd}}+\frac32s_{ab}^2\right)
    \end{equation}
and
\begin{align}
    R^{(2)}_{5:1B}(a^+,b^+,c^+,d^+,e^+)&=2i\epsilon\left(a,b,c,d\right)\Big(\text{C}_{\text{PT}}(a,b,e,c,d)+\text{C}_{\text{PT}}(a,d,b,c,e)\notag\\
    &\  +\text{C}_{\text{PT}}(b,c,a,d,e)+\text{C}_{\text{PT}}(a,b,d,e,c)+\text{C}_{\text{PT}}(a,c,d,b,e)\Big)
\end{align}
where 
\begin{align}
    \text{C}_{\text{PT}}(a,b,c,d,e)=\frac{1}{\spa{a}.b\spa{b}.c\spa{c}.d\spa{d}.e\spa{e}.a}\,.
\end{align}
These expressions are valid for both $U(N_c)$ and $SU(N_c)$ gauge groups and are remarkably compact.

We note that there are double poles at leading and sub-leading in color, but not at sub-sub-leading. 
As $R_{4:1B}^{(2)}$ vanishes~\cite{Bern:2002tk} the poles 
in $R_{5:1B}^{(2)}$ do not correspond to tree:two-loop factorisations, instead they arise from contributions of the type shown in
fig.\ref{fig:axialg} where the corresponding current has no pole in $s_{ab}$.

\section{Conclusions}
Computing perturbative gauge theory amplitudes to high orders is an important but difficult task. 
In this article, we have demonstrated how the full color all-plus five-point amplitude may be computed in simple forms.  
We have computed all the color components directly and only used color relations between them as checks.  In passing, 
we have given simple all-$n$ expressions for the one-loop 
subleading in color amplitudes and presented the $n$-point IR divergences in a color basis approach.  Our methodology
obtains these results without the need to determine two-loop non-planar integrals.

\section{Acknowledgements}
DCD was supported by STFC grant ST/L000369/1. JMWS was 
supported by STFC grant ST/S505778/1. JHG was 
supported by the College
of Science (CoS) Doctoral Training Centre (DTC) at 
Swansea University.

\appendix

\section{Infra-Red Divergences} 
\label{ap:IR}

The singular behaviour of two-loop gluon scattering amplitudes is known from a general analysis~\cite{Catani:1998bh}.
The leading IR singularity for the $n$-point two-loop amplitude is~\cite{Bern:2000dn}
\begin{equation}
-\frac{s_{ab}^{-\epsilon}}{\epsilon^2} f^{aij}f^{bik} \times {\cal A}_n^{(1)}(j,k,\cdots, n )
\end{equation}
where ${\cal A}_n^{(1)}$ is the full-color one-loop amplitude.  
We wish to disentangle this simple equation into the color-ordered partial amplitudes. 
It will be convenient to use a more list based notation for the  partial amplitudes where
we use
\begin{equation}
A_n^{(l)}(S) = A_n^{(l)}(\{ a_1, a_2 , \cdots a_n \}) \equiv  A_{n:1}^{(l)}(a_1, a_2 , \cdots a_n ) \,,
\end{equation}
$A_n^{(l)}(S_1;S_2)$ for $A^{(l)}_{n:r}$ and
$A_n^{(l)}(S_1;S_2;S_3)$ for $A^{(l)}_{n:s,t}$. 

First we define
\begin{equation}
I_{i,j} \equiv -\frac{(s_{ij})^{-\epsilon}}{\epsilon^2}
\end{equation}
and we have for a list $S=\{a_1,a_2,a_3,\cdots, a_r \}$,
\begin{equation}
I_r[S] =\sum_{i=1}^r I_{a_i,a_{i+1}} 
\end{equation}
where the term $I_{a_r,a_{r+1}}\equiv I_{a_r,a_{1}}$ is included in the sum.   We also define $I_j[S_1,S_2]$ and $I_k[S_1,S_2]$, 
\begin{eqnarray}
I_j[ S_1,S_2]=I_j [ \{a_1,a_2\cdots a_r \}, \{b_1,b_2,\cdots b_s\} ]  \equiv
\left( I_{a_1,a_r}+I_{b_1,b_s} -I_{a_1,b_1} -I_{a_r,b_s}  \right)\,,
\notag \\
I_k[ S_1,S_2]=I_k [ \{a_1,a_2\cdots a_r \}, \{b_1,b_2,\cdots b_s\} ]  \equiv
\left( I_{a_1,b_s}+I_{b_1,a_r} -I_{a_1,b_1} -I_{a_r,b_s}  \right)
\end{eqnarray}
giving
\begin{equation}
I_r[S_1\oplus S_2] =I_r[S_1]+I_r[S_2]+I_k[S_1,S_2]-I_j[S_1,S_2]
\end{equation}
where $\{a_1\cdots a_r \} \oplus \{b_1\cdots b_s\}=\{a_1\cdots a_r,b_1\cdots b_s\}$. 
In this language the leading and subleading IR singularities at one-loop are
\begin{eqnarray}
A_{n}^{(1)} (S) &=& A_{n}^{(0)} (S) \times
   I_r[S ] \,,
\notag \\
 A_{n}^{(1)} (S_1 ; S_2 ) &=&  \sum_{S'_1\in C(S_1)} \sum_{S'_2 \in C(S_2)} A_{n}^{(0)} (S'_1 \oplus S'_2 )\times I_j[ S'_1,S'_2]  \,.
\end{eqnarray}
The set $C(S)$ is the set of cyclic permutations of $S$. 

At two-loops, we have
\begin{align}
A_{n}^{(2)} (S) & = A_{n}^{(1)} (S) 
\times I_r[S]  \,, 
\notag\\
A_{n}^{(2)} (S_1 ; S_2 ) &=  A_{n}^{(1)} (S_1 ; S_2 ) \times  \left( I_r[S_1]+I_r[S_2] \right) 
\notag \\
& \qquad +  \sum_{S'_1\in C(S_1)} \sum_{S'_2\in C(S_2)} A_{n}^{(1)} (S'_1 \oplus S'_2 )\times I_j[ S'_1,S'_2]  \,,  
\notag \\
A_{n}^{(2)} (S_1 ; S_2; S_3 ) &=    \sum_{S'_2\in C(S_2)} \sum_{S'_3\in C(S_3)} A_{n}^{(1)} (S_1; S'_2 \oplus S'_3 )\times I_j[ S'_2,S'_3]  
\notag \\
&\qquad  +  \sum_{S'_1\in C(S_1)} \sum_{S'_3\in C(S_3)} A_{n}^{(1)} (S_2; S'_1 \oplus S'_3 )\times I_j[ S'_1,S'_3]  
\notag \\
& \qquad +  \sum_{S'_1\in C(S_1)} \sum_{S'_2\in C(S_2)} A_{n}^{(1)} (S_3; S'_1\oplus S'_2 )\times  I_j[ S'_1,S'_2]   \,, 
\notag \\
A_{n,B}^{(2)}  (S) &= \sum_{U(S)}  A_{n}^{(1)}  (S'_1;S'_2) 
\times  I_k[S'_1,S'_2]    \,, 
\end{align}
where $U(S)$ is the set of all distinct pairs of lists satisfying $S'_1\oplus S'_2 \in C(S)$  where the size of $S'_i$ is greater than one. For example
\begin{align}
U(\{1,2,3,4,5\})= \biggl\{ &
(\{1,2\},\{3,4,5\} ),  
(\{2,3\},\{4,5,1\} ), 
(\{3,4\},\{5,1,2\} ), 
\notag \\ &
(\{4,5\},\{1,2,3\} ), 
(\{5,1\},\{2,3,4\} ) 
\biggr\}\,.
\end{align}

\section{Obtaining the adjacent current}
\label{app:reco}
We build the rational part of the full color five-point amplitude recursively, using augmented recursion to determine the sub-leading poles arising 
in the one-loop:one-loop factorisations. For this we need an approximation to the doubly massive current $\tau_5^{(1)}(\alpha,\beta,c,d,e)$ shown in
fig.\ref{fig:axialg}.
As we are only interested in the residue on the $s_{ab}\to 0$ pole, we do not need the exact current, just one that satisfies the conditions:
\begin{enumerate}
    \item[(C1)] The current contains the leading singularity as $s_{\alpha\beta}\to0$ with $\alpha^2,\beta^2\neq0$,
    \item[(C2)] The current is the one-loop, single-minus amplitude in the on-shell limit $\alpha^2,\beta^2\to0$, $s_{\alpha\beta}\neq0$.
\end{enumerate} 

Condition (C2) requires the current $\tau^{(1)}_{5:\lambda}$ to reproduce the full partial amplitude $A^{(1)}_{5:\lambda}$ in the $\alpha^2\to0$, $\beta^2\to0$ 
limit and so  the current should have the same color decomposition as the one-loop amplitude (\ref{eq:oneloopcolordeco}).
We can use \eqref{eq_COP} to relate any of the sub-leading currents to sums of the leading  in color currents $\tau^{(1)}_{5:1}$. 
The cyclic and flip symmetries inherited from  $A^{(1)}_{5:1}$ mean that any of the $\tau^{(1)}_{5:\lambda}$ can be related 
to $\tau^{(1)}_{5:1}(\alpha,\beta,c,d,e)$ and $\tau^{(1)}_{5:1}(\alpha,c,\beta,d,e)$ up to permutations of the legs $\{c,d,e\}$.

To build the current we start with the one-loop, five-point, single-minus partial amplitude
\begin{align}
    A^{(1)}_{5:1}(\alpha^-,\beta^+,c^+,d^+,e^+)= \sum_{j=i,ii,iii}A^{(1)}_{5:1j}\left(\alpha^-,\beta^+,c^+,d^+,e^+\right)
    \label{samp}
\end{align}
where
\begin{align*}
    A^{(1)}_{5:1i}\left(\alpha^-,\beta^+,c^+,d^+,e^+\right)&=\frac{i}{3}\frac{1}{\spa{c}.d^2}
    \frac{\spa{c}.e\spa{\alpha}.d^3\spb{d}.e}{\spa{\alpha}.{\beta}\spa{d}.e^2\spa{\beta}.c},\\
       A^{(1)}_{5:1ii}\left(\alpha^-,\beta^+,c^+,d^+,e^+\right)&=-\frac{i}{3}\frac{1}{\spa{c}.d^2}
   \frac{\spb{\beta}.e^3}{\spb{\alpha}.{\beta}\spb{e}.{\alpha}} \,
  \intertext{and}
    A^{(1)}_{5:1iii}\left(\alpha^-,\beta^+,c^+,d^+,e^+\right)&= \frac{i}{3}\frac{1}{\spa{c}.d^2}
  \frac{\spa{\alpha}.c^3 \spa{\beta}.d \spb{\beta}.c}{\spa{d}.e\spa{\alpha}.e \spa{\beta}.c^2}\,.\\
\end{align*}

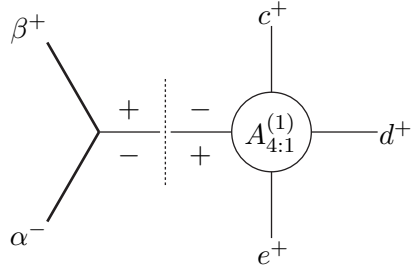
\begin{figure}[H]
\centering
\begin{picture}(170,150) 
	\Line(20,50)(43,50)
	\DashLine(45,70)(45,30){1}
	\Text(31.5, 55.)[b]{$+$}
	\Text(31.5, 45.)[t]{$-$}
	\Text(58.5, 55.)[b]{$-$}
	\Text(58.5, 45.)[t]{$+$}
	\Line(47,50)(70,50)
	\CCirc(85,50){15}{Black}{White}
	\Line(85,65)(85,90) \Text(86,91)[b]{$c^+$}
	\Line(100,50)(125,50) \Text(126,50)[l]{$d^+$}
	\Line(85,35)(85,10) \Text(86,9)[t]{$e^+$}
	\SetWidth{1.0}
	\Line(20,50)(0.5,16.225)
	\Line(20,50)(0.5,83.775)
	\Text(0.5,16.225)[tr]{$\alpha^-$}
	\Text(0.5,83.775)[br]{$\beta^+$}
	\Text(85,50)[c]{$A^{(1)}_{4:1}$}
\end{picture}
\caption{Factorisations of the current on the $s_{\alpha\beta}\to0$ pole.}
\label{fig:facchans}
\end{figure}

Condition (C1) requires our approximation to the current to reproduce the correct leading singularities as $s_{\alpha\beta}\to0$, 
the sources of these are depicted in fig.\ref{fig:facchans}~\cite{Dunbar:2016aux}. We determine these within the axial gauge
formalism. The two channels give

\begin{align}
       \mathcal{F}^{\;cde}_{dp}&\equiv \frac{\spb{\beta}.k\spa{\alpha}.q^2}{\spa{\beta}.q\spa{k}.q}\frac1{s_{\alpha\beta}}A^{(1)}_{4:1}(k^-,c^+,d^+,e^+)
        =\frac{i}3\frac{\spa{\alpha}.q^2}{\spa{\beta}.q^2}\frac{\la q|\alpha\beta|q\ra}{s_{\alpha\beta}}
          \frac{\spa{e}.c\spb{c}.e^3}{\spa{c}.d\spa{d}.e[e|P_{\alpha\beta}|q\ra[c|P_{\alpha\beta}|q\ra}
     \notag \\
     \intertext{and}
     \mathcal{F}^{\;cde}_{sb}&\equiv\frac{\spa{\alpha}.k\spb{\beta}.q^2}{\spb{\alpha}.q\spb{k}.q} 
     \frac{1}{s_{\alpha\beta}}A^{(1)}_{4:1}(k^+,c^+,d^+,e^+)=-\frac{i}{3}\frac{\spa{\alpha}.k\spb{\beta}.q^2}{\spb{\alpha}.q\spb{k}.q}\frac{1}{s_{\alpha\beta}}\frac{\spb{e}.k^2}{\spa{c}.d^2}
 \end{align}
where $k=\alpha+\beta=-c-d-e$ which is null on the pole.

Using the identity
\begin{equation}
    \begin{split}
        \frac1{\spa{\alpha}.{\beta}\spa{\beta}.c}=\frac1{\spa{\alpha}.q\spa{\beta}.q^2}\left(
        \frac{\la q|\alpha\beta|q\ra[q|P_{\alpha\beta}|q\ra}{s_{\alpha\beta}[q|P_{\alpha\beta}|c\ra}
        +\frac{\spa{q}.{\beta}\spa{q}.c[q|\alpha|q\ra}{\spa{\beta}.c[q|P_{\alpha\beta}|c\ra}\right)
    \end{split}
\end{equation}
and the expansion
\begin{align}
    \frac{[\beta|P^{\flat}_{\alpha\beta}|d\ra}{[\beta|P_{\alpha\beta}|q\ra}=\frac{[q|P_{\alpha\beta}|d\ra}
    {[q|P_{\alpha\beta}|q\ra}+s_{\alpha\beta}\frac{\spa{q}.d\spb{\beta}.q}{[\beta|P_{\alpha\beta}|q\ra[q|P_{\alpha\beta}|q\ra}+\mathcal{O}(s_{\alpha\beta}^2)
\end{align}
we find
\begin{equation}
    A^{(1)}_{5:1i}=   \mathcal{F}^{\;cde}_{dp}\left[1+s_{\alpha\beta}\left(
    \frac{\spb{q}.e}{\spb{c}.e[q|P_{\alpha\beta}|c\ra}+\frac{[c|q|d\ra}{[q|P_{\alpha\beta}|q\ra[c|e|d\ra}
    +\frac{[e|q|d\ra}{[q|P_{\alpha\beta}|q\ra[e|c|d\ra}\right)+\mathcal{O}(s_{\alpha\beta}^2)\right]. 
    \label{eq:dpbitty}
\end{equation}
We see that $A^{(1)}_{5:1i}$ generates the correct singularity as $\spa{\alpha}.\beta\to 0$. This terms generates the double pole when integrated and the
form in (\ref{eq:dpbitty}) explicitly exposes the subleading contribution.  

The $\mathcal{F}^{\;cde}_{sb}$ factorisation arises when $\spb{\alpha}.\beta\to 0$. This we obtain from $A^{(1)}_{5:1ii}$. 
 Using,
\begin{align}
    k^{\flat}=k-\frac{k^2}{2k.q}q=\alpha^{\flat}+\beta^{\flat}+\delta q,
\end{align}
where 
\begin{align}
    \delta=\frac{\alpha^2}{2\alpha.q}+\frac{\beta^2}{2\beta.q}-\frac{s_{\alpha\beta}}{2k.q},
\end{align}
we have
\begin{equation}
        \mathcal{F}^{\;cde}_{sb}=
        \frac{i}3\frac1{s_{\alpha\beta}}\left[\frac{\spb{e}.{\beta}^2\spb{\beta}.q\spa{\beta}.{\alpha}}{\spb{\alpha}.q\spa{c}.d^2}+\delta[e|q|\alpha\ra
        \frac{(\spb{e}.{\beta}\spb{\beta}.q\spb{k}.q+\spb{\beta}.q^2\spb{e}.k)}{\spb{\alpha}.q\spb{k}.q\spa{c}.d^2}\right].
    \label{factor2}
\end{equation}
Now $A^{(1)}_{5:1ii}$ can be rewritten as
\begin{equation}
      A^{(1)}_{5:1ii}  
        =-\frac{i}3\frac1{\spa{c}.d^2}\frac{\spb{\beta}.e^2\spb{q}.e}{\spb{e}.{\alpha}\spb{\alpha}.q}
        -\frac{i}3\frac1{\spa{c}.d^2}\frac{\spb{\beta}.e^2\spb{q}.{\beta}}{\spb{\alpha}.{\beta}\spb{\alpha}.q}
    \label{aiii}
\end{equation}
and noting that
\begin{equation}
    \begin{split}
        \frac{\spa{\beta}.{\alpha}}{s_{\alpha\beta}}-\frac1{\spb{\alpha}.{\beta}}
        =\frac{\spa{\beta}.{\alpha}\spb{\alpha}.{\beta}-s_{\alpha\beta}}{s_{\alpha\beta}\spb{\alpha}.{\beta}}
        =\frac{(\alpha^{\flat}+\beta^{\flat})^2-s_{\alpha\beta}}{s_{\alpha\beta}\spb{\alpha}.{\beta}}
        =-\left(\frac{\alpha^2}{2\alpha.q}+\frac{\beta^2}{2\beta.q}\right)\frac{2k.q}{s_{\alpha\beta}\spb{\alpha}.{\beta}}\,,
    \end{split}
    \label{offshell}
\end{equation}
we see that $A^{(1)}_{5:1ii}$ has the form $\mathcal{F}^{\;cde}_{sb}+\Delta_{\alpha^2}+\Delta_{\beta^2}+\Delta_{s_{\alpha\beta}}$ as $\spb{\alpha}.\beta\to 0$, 
where $\Delta_{\alpha^2}$ is proportional to $\alpha^2/s_{\alpha\beta}$ etc..  As $\Delta_{s_{\alpha\beta}}$  does not contribute on the pole, 
we don't have to replicate it in the current and therefore include a contribution to the current of the form $A^{(1)}_{5:1ii}-\Delta_{\alpha^2}-\Delta_{\beta^2}$  
to satisfy condition (C1). This does not compromise condition (C2). Upon integration the $\alpha^2$ and $\beta^2$ factors in $\Delta_{\alpha^2}$ and
$\Delta_{\beta^2}$ generate $s_{ab}$ factors which cancel the pole. We therefore do not require these forms explicitly.  For the purposes of integration
it is convenient to express the term with the $\spb{\alpha}.\beta$ pole in terms of $\mathcal{F}^{\;cde}_{sb}$. To maintain condition (C2) we must retain
$\Delta_{s_{\alpha\beta}}$.

The remaining piece of the one-loop amplitude, $A^{(1)}_{5:1iii}$, contains no poles as  $\spa{\alpha}.\beta\to 0$ or
$\spb{\alpha}.\beta\to 0$ and we can simplify it using

\begin{align}
    \frac{\spa{X}.{\alpha}}{\spa{Y}.{\alpha}} =  \frac{\spa{X}.{\alpha}}{\spa{Y}.{\alpha}} 
    \frac{\spa{Y}.a}{\spa{Y}.a}=\frac{\spa{X}.a}{\spa{Y}.a}+\mathcal{O}(\spa{\alpha}.a)
\end{align}
as terms of $\mathcal{O}(\spa{\alpha}.a)$ do not ultimately contribute to the residue. 

The adjacent leading current is then
\begin{equation}
    \begin{split}
       &\tau^{(1)}_{5:1}(\alpha^-,\beta^+,c^+,d^+,e^+) =
       \mathcal{F}^{\;cde}_{dp}\left[1+s_{\alpha\beta}\left(\frac{\spb{q}.e}{\spb{c}.e[q|P_{\alpha\beta}|c\ra}
       +\frac{[c|q|d\ra}{[q|P_{\alpha\beta}|q\ra[c|e|d\ra}+\frac{[e|q|d\ra}{[q|P_{\alpha\beta}|q\ra[e|c|d\ra}\right)\right]\\
      &+\frac{i}{3\spa{c}.d^2}\frac{\spa{\alpha}.q^2}{\spa{\beta}.q^2}\left[\frac{\spa{a}.c[c|\beta|d\ra}{\spa{d}.e\spa{e}.a}
       +\frac{\spa{c}.e\spb{d}.e}{\spa{d}.e^2}\left(\frac{[q|P_{\alpha\beta}|d\ra^3}{[q|P_{\alpha\beta}|q\ra^3}
       \frac{\spa{q}.c\spa{q}.a[q|\alpha|q\ra}{\spa{a}.c[q|P_{\alpha\beta}|c\ra}
       -3\frac{\spa{q}.d[q|P_{\alpha\beta}|d\ra^2[q|\beta|q\ra}{[q|P_{\alpha\beta}|q\ra^2[q|P_{\alpha\beta}|c\ra}\right)\right]\\
       &+\mathcal{F}^{\;cde}_{sb}+\frac{i}{3\spa{c}.d^2}
       \left(-\frac{\spb{\beta}.e^2\spb{q}.e}{\spb{e}.{\alpha}\spb{\alpha}.q}
       +[e|q|\alpha\ra\frac{(\spb{e}.{\beta}\spb{\beta}.q\spb{k}.q+\spb{\beta}.q^2\spb{e}.k)}{\spb{\alpha}.q\spb{k}.q2k.q}\right)
       +\mathcal{O}(\spa{\alpha}.\beta)+\mathcal{O}(\alpha^2)+\mathcal{O}(\beta^2)\,.
    \end{split}
\end{equation}



\bibliography{TwoLoop}{}

\bibliographystyle{ieeetr}

\end{document}